\documentclass[rsi,amsmath,amssymb,preprint]{revtex4-1}
\usepackage{graphicx}		% Include figure files
\usepackage{dcolumn}		% Align table columns on decimal point
\usepackage{bm}			% bold math

\def\wisk#1{\ifmmode{#1}\else{$#1$}\fi}

\def\deg    {\wisk{^\circ}}

\def\apj {The Astrophysical Journal}
\def\apjs {The Astrophysical Journal Supplement Series}
\def\prl {Physical Review Letters}

\begin{document}

\title{Superfluid Liquid Helium Control for the \\
Primordial Inflation Polarization Explorer Balloon Payload}

% Author list 
\author{A. Kogut}
  \affiliation{Code 665, Goddard Space Flight Center, Greenbelt MD 20771 USA}
  \email{alan.j.kogut@nasa.gov}
\author{T. Essinger-Hileman}
  \affiliation{Code 665, Goddard Space Flight Center, Greenbelt MD 20771 USA}
\author{D. Fixsen}
  \affiliation{Department of Astronomy, University of Maryland, College Park MD 20740 USA}
  \altaffiliation{Code 665, Goddard Space Flight Center, Greenbelt MD 20771 USA}
\author{L. Lowe}
  \affiliation{Sigma Space Corp, Lanham MD 20706 USA}
  \altaffiliation{Code 665, Goddard Space Flight Center, Greenbelt MD 20771 USA}
\author{P.Mirel}
  \affiliation{Sigma Space Corp, Lanham MD 20706 USA}
  \altaffiliation{Code 665, Goddard Space Flight Center, Greenbelt MD 20771 USA}
\author{E. Switzer}
  \affiliation{Code 665, Goddard Space Flight Center, Greenbelt MD 20771 USA}
\author{E. Wollack}
  \affiliation{Code 665, Goddard Space Flight Center, Greenbelt MD 20771 USA}

% \date{\today}
% \date{April 15, 2021}

% ----------- Abstract -----------
\begin{abstract}
The Primordial Inflation Polarization Explorer (PIPER)
is a stratospheric balloon payload
to measure polarization of the cosmic microwave background.
Twin telescopes mounted within an open-aperture bucket dewar
couple the sky to bolometric detector arrays.
We reduce detector loading and photon noise
by cooling the entire optical chain to 1.7~K or colder.
A set of fountain-effect pumps
sprays superfluid liquid helium
onto each optical surface,
producing helium flows of
50--100~cm$^3$~s$^{-1}$
at heights up to 200 cm above the liquid level. 
We describe the fountain-effect pumps
and the cryogenic  performance of the PIPER payload
during two flights in 2017 and 2019.
\end{abstract}

\maketitle

% --------------------- Begin main text ---------------------

\section{Introduction}
Linear polarization of the cosmic microwave background (CMB)
provides a window into the physics of the early universe.
Gravitational waves created during an inflationary epoch
interact with CMB photons at much later times 
to impart a distinctive pattern 
in linear polarization.
For the simplest (single-field) inflation models,
the amplitude of the polarized signal 
depends on the energy scale of inflation as
%
%
%
%
%	\vspace{-2mm}
\begin{equation}
E = 1.06 \times 10^{16} 
~(r / 0.01)^{1/4} 
~{\rm GeV} 
\label{inflation_eq}
%	\vspace{-2mm}
\end{equation}
where 
$r$ is the power ratio of tensor
(gravitational wave) to scalar (density) fluctuations
sourced during inflation
\cite{lyth/riotto:1999}.
In most large-field models of inflation,
$r$ is predicted to be of order 0.01,
corresponding to energy near the Grand Unified Theory scale, $10^{16}$ GeV.

Current upper limits $r < 0.06$
\cite{bicep2xkeck_2018}
correspond to fluctuations in surface brightness
$\Delta T < 80$~nK
on angular scales of a few degrees 
(in units of CMB thermodynamic temperature).
At millimeter wavelengths where CMB emission peaks,
the sensitivity of bolometric detectors
has reached the point 
where a dominant contribution to the system noise is
photon noise from the random arrival of incident photons.
The spectral density of photon noise
in a single linear polarization 
from a source with emissivity $\epsilon$
at physical temperature $T$
is given by
\begin{equation}
{\rm NEP}^2_{\rm photon} = {2A\Omega \over c^2} {(kT)^5\over h^3}
	\int \alpha \epsilon f
	  ~\frac{x^4}{e^x-1} 
	  ~\left( 1 + \frac{\alpha \epsilon f}{e^x-1} \right) ~dx ,
\label{photon_nep_eq}
\end{equation}
where
$x=h\nu/kT$,
$h$ is the Planck constant,
$k$ is the Boltzmann constant,
$c$ is the speed of light,
$\nu$ is the observing frequency,
$A$ is the detector area,
$\Omega$ is the solid angle,
$\alpha$ is the detector absorptivity,
and
$f$ is the power transmission through the optics
\cite{mather:1982}.
The noise $\delta P$ 
at the detector
may be referred to the surface brightness on the sky,
\vspace{-2mm}
\begin{equation}
\delta I_\nu = \frac{ \delta P }
		      { A\Omega ~\Delta \nu ~(\alpha \epsilon f) }
\label{i_noise}
\end{equation}
where 
$\Delta \nu$ is the observing bandwidth.
Even an ideal noiseless detector
($f=1$ and $\alpha=1$)
with a 20\% fractional bandwidth at 200 GHz
will have irreducible noise 
from CMB photons
of order 20~$\mu$K within each 1-second integration.
Realistic instruments
($f=0.3$ and $\alpha=0.5$)
perform correspondingly worse.
Emission from the atmosphere 
and optical elements skyward of the detector
further degrade sensitivity.
Detecting signals at the few-nK level
requires a combination of long integration times
and multiple independent detectors.
Observations at altitudes above 30 km
minimize photon noise from the atmosphere,
improving per-detector sensitivity
by a factor of ten or more 
compared to the best ground-based sites.
Additional reductions in photon noise
of factors 1.5--4 
can be achieved by cooling optical elements
to cryogenic temperatures.

%--------------------------------------------------------------------------
% Figure 1: PIPER schematic
%--------------------------------------------------------------------------
\begin{figure}[b]
\centerline{
\includegraphics[height=3.5in]{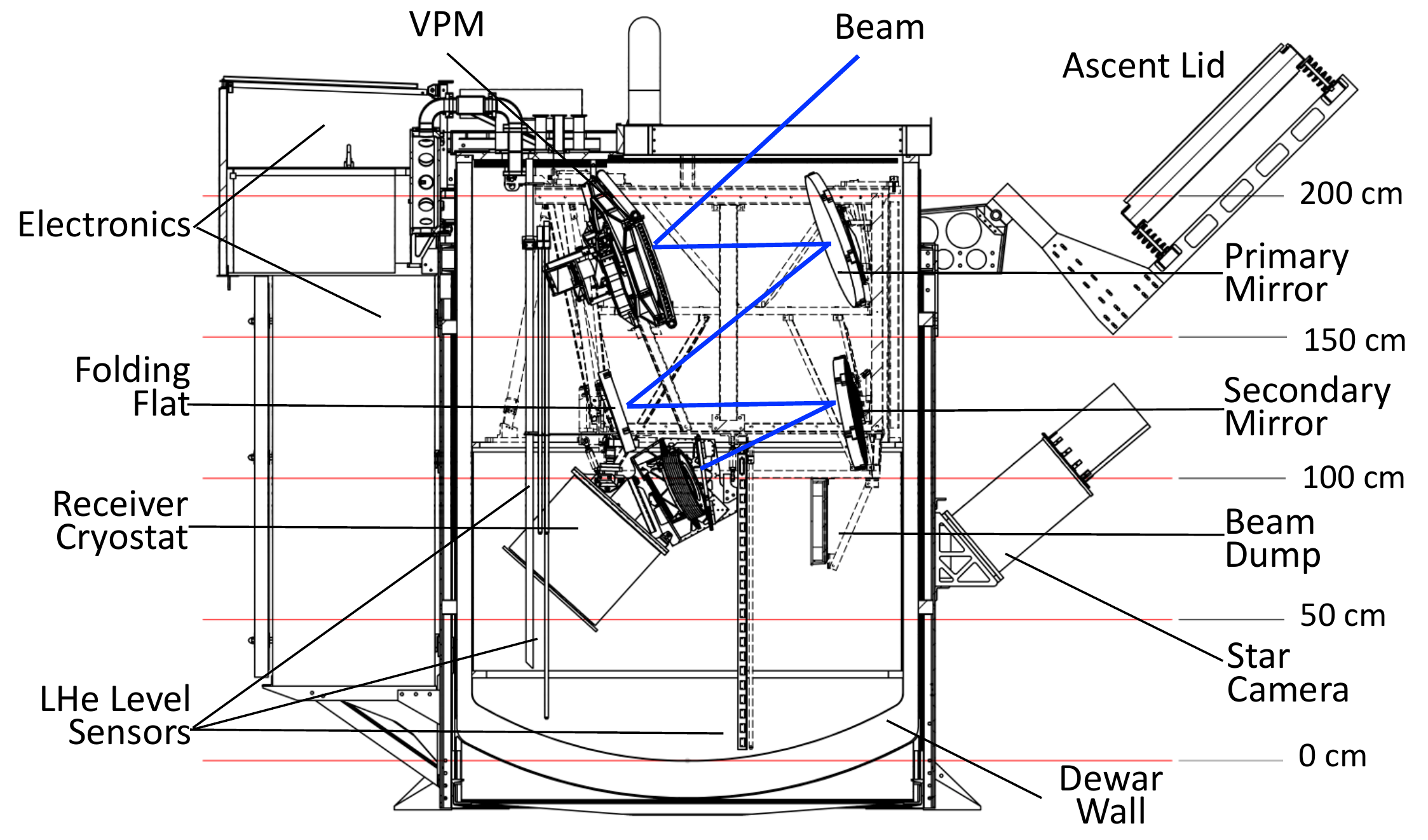} }
\caption{
Schematic drawing of the PIPER payload.
All optical elements are mounted inside a large liquid helium
bucket dewar.
Superfluid pumps spray liquid helium
to maintain each optical surface
at temperature 1.7~K.
Red lines indicate the height in cm
above the dewar bottom.
}
\label{piper_schematic}
\end{figure}
%--------------------------------------------------------------------------

The Primordial Inflation Polarization Explorer (PIPER)
is a stratospheric balloon payload
to measure polarization of the CMB and diffuse astrophysical foregrounds
in four frequency bands between 200 and 600 GHz\cite{
piper_spie_2012,
%	piper_spie_2014,
piper_spie_2018}.
Figure \ref{piper_schematic} shows the payload.
It consists of two co-pointed telescopes mounted within 
an open liquid helium bucket dewar.
An ascent lid covers the dewar aperture during ground operations
and opens at float altitude to allow the telescopes
to view the sky.
A variable-delay polarization modulator (VPM) on each telescope
modulates linear and circular polarization 
to isolate the polarized signal 
while rejecting the much brighter unpolarized emission\cite{
vpm_2012,
vpm_2014}.
A pair of $32 \times 40$ element detector arrays 
operating at 100 mK within a vacuum receiver cryostat
provide background-limited sensitivity\cite{piper_receiver_2019}.
A continuous adiabatic demagnetization refrigerator
provides sub-K cooling for the detectors;
its design and operation is described in \cite{piper_receiver_2019}.
Electronics racks mounted on the dewar exterior frame
provide detector readout,
housekeeping,
avionics,
and telemetry.
The entire payload rotates in azimuth
about the dewar vertical axis;
gyroscopes, 
magnetometers
and a star camera 
provide pointing information
for real-time control
and post-flight pointing reconstruction.

%--------------------------------------------------------------------------
% Figure 2: Photon NEP
%--------------------------------------------------------------------------
\begin{figure}[b]
\centerline{
\includegraphics[height=3.5in]{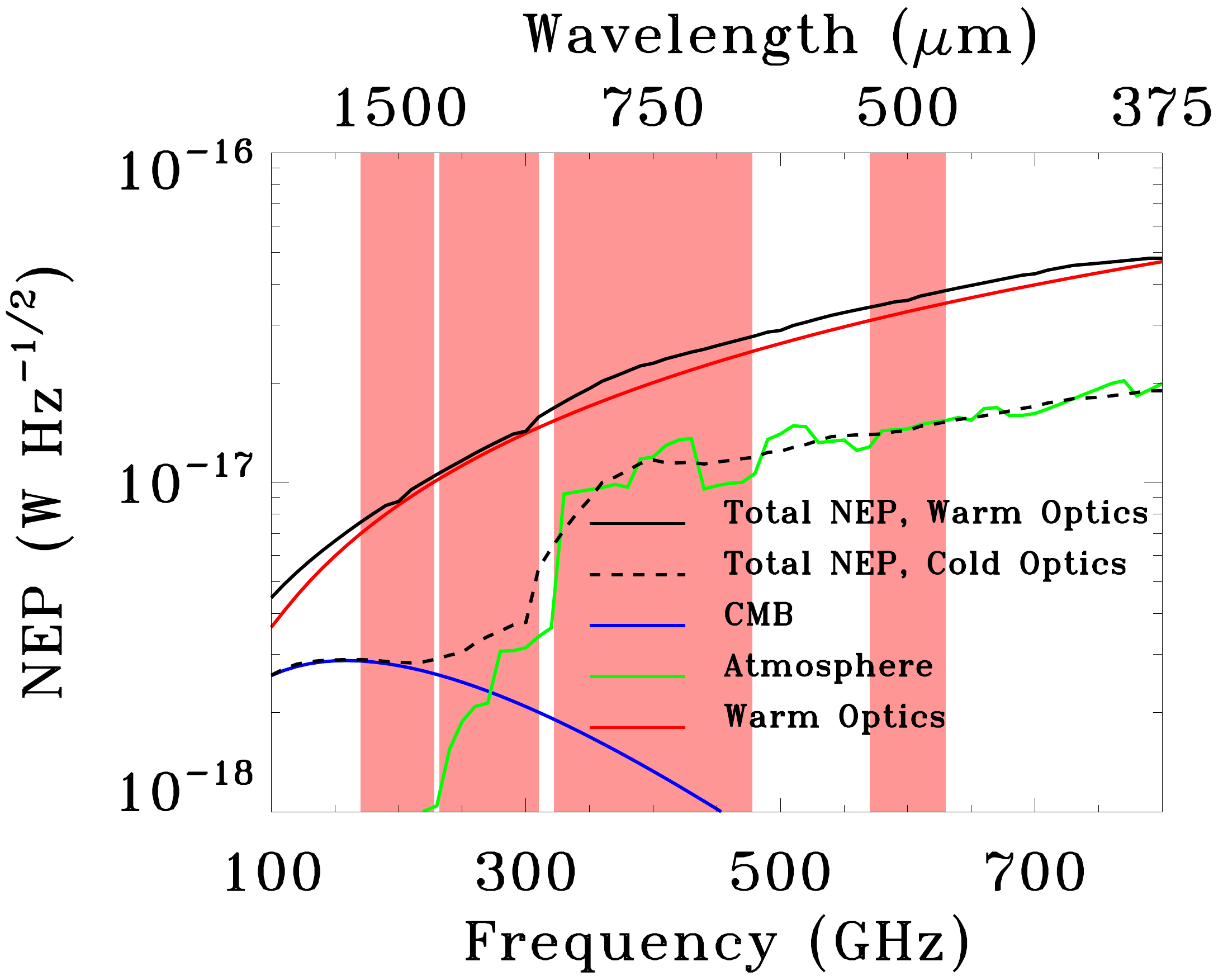} }
\caption{
Noise Equivalent Power from photon statistics
are shown for the CMB,
atmosphere,
and PIPER reflective optics
for cases with the optics at 250~K vs 1.7~K.
Pink bands show the PIPER frequency bands.
Cooling the optics below 10~K
improves the system NEP by a factor of 3,
equivalent to increasing the detector count by a factor of 9.}
\label{nep_fig}
\end{figure}
%--------------------------------------------------------------------------

PIPER's optical design
has 4 reflective surfaces
between the sky and the receiver:
the VPM, primary mirror, folding flat, and secondary mirror\cite{
eimer/etal:2010}.
Figure \ref{nep_fig}
shows the photon noise
from the CMB, atmosphere,
and the reflective optics.
Maintaining these elements at temperatures below 10K
effectively eliminates their contribution to the photon noise,
improving the noise equivalent power (NEP)
by a factor of 3 compared to a comparable system
with ambient temperature (250~K) optics.
Achieving a comparable improvement
with warm optics
would require increasing the number of detectors
by a factor of 9.

PIPER's azimuth scan requires the beams to exit the dewar
at an angle of 55\deg~from the zenith.
The open bucket dewar cannot be tipped more than a few degrees
from vertical without compromising its performance;
consequently,
the VPM and primary mirrors
must be located near the top of the dewar.
Cooling from the efflux of boiloff gas alone
is insufficient
to maintain these elements at temperatures below 10~K.
Maintaining the optics below 10~K
requires a constant flow of liquid helium
delivered directly to the PIPER mirrors.

PIPER operates at altitudes above 30 km.
At the resulting ambient pressure of 1~kPa or less,
liquid helium within the open bucket dewar 
boils at temperature below 1.7~K,
well beneath the 2.2~K superfluid transition temperature.
Liquid helium at temperatures below the $\lambda$ point
may be modeled as two weakly interacting fluids:
a normal component 
and a superfluid component with zero viscosity and zero entropy.
The low temperature
and zero viscosity of the superfluid
make standard mechanical pumps impractical.
Superfluid pumps employing the thermomechanical effect
provide a robust method 
to move macroscopic quantities of superfluid helium
throughout the PIPER dewar
without requiring moving parts
and without injecting 
vibration or electromagnetic noise
into the PIPER payload.

The thermomechanical effect in liquid helium
was first described by
\cite{allen/jones:1938}
with a theoretical explanation by
\cite{london:1938, london:1939}.
Superfluid pumps have been used
to cool optical elements
in previous balloon flights
\cite{woody/richards:1981,
peterson/etal:1985,
bernstein/etal:1990,
kogut/etal:2004,
singal/etal:2011}
and
have been demonstrated 
in space missions at zero gravity\cite{
dipirro/castles:1986,
dipirro2016}.
PIPER differs from these in several important aspects.
Balloon-borne open-aperture optics cooled below 10~K
provide a platform to enable
millimeter through mid-infrared observations
with near-space optical loads.
The physical scale of the PIPER optics
(50~cm diameter mirrors mounted 200~cm above the dewar floor)
significantly exceeds the
6~cm diameter and 58~cm height
of early balloon experiments\cite{woody/richards:1981,
peterson/etal:1985,
bernstein/etal:1990},
requiring a
substantially greater cooling flow.
The ARCADE-2 mission
demonstrated 30~cm open-aperture at frequencies 3--90~GHz\cite{
singal/etal:2011,
fixsen/etal:2011}.
PIPER extends comparable open-aperture operation
to larger apertures
and frequencies 200--600~GHz\cite{piper_spie_2018}.
The EXCLAIM balloon mission\cite{cataldo/etal:2021}
plans to use a similar approach
for 1~m optics at far-infrared wavelengths,
while the BOBCAT program\cite{kogut/etal:2021}
seeks to fly cold 3--4~meter optics.
We describe the superfluid pumps
for the PIPER balloon payload
and  discuss the thermal performance and cryogenic managment
of the PIPER payload
for two flights in 2017 and 2019.
These pumps 
were originally developed for the ARCADE balloon payload\cite{
kogut/etal:2004,
singal/etal:2011},
but have not previously been described.
The in-flight performance of 
the PIPER superfluid pumps
and the
resulting temperature profiles within the open bucket dewar
provide a benchmark for development of future balloon missions.

%--------------------------------------------------------------------------
% Figure 3: Pump schematic
%--------------------------------------------------------------------------
\begin{figure}[b]
\centerline{
\includegraphics[height=3.5in]{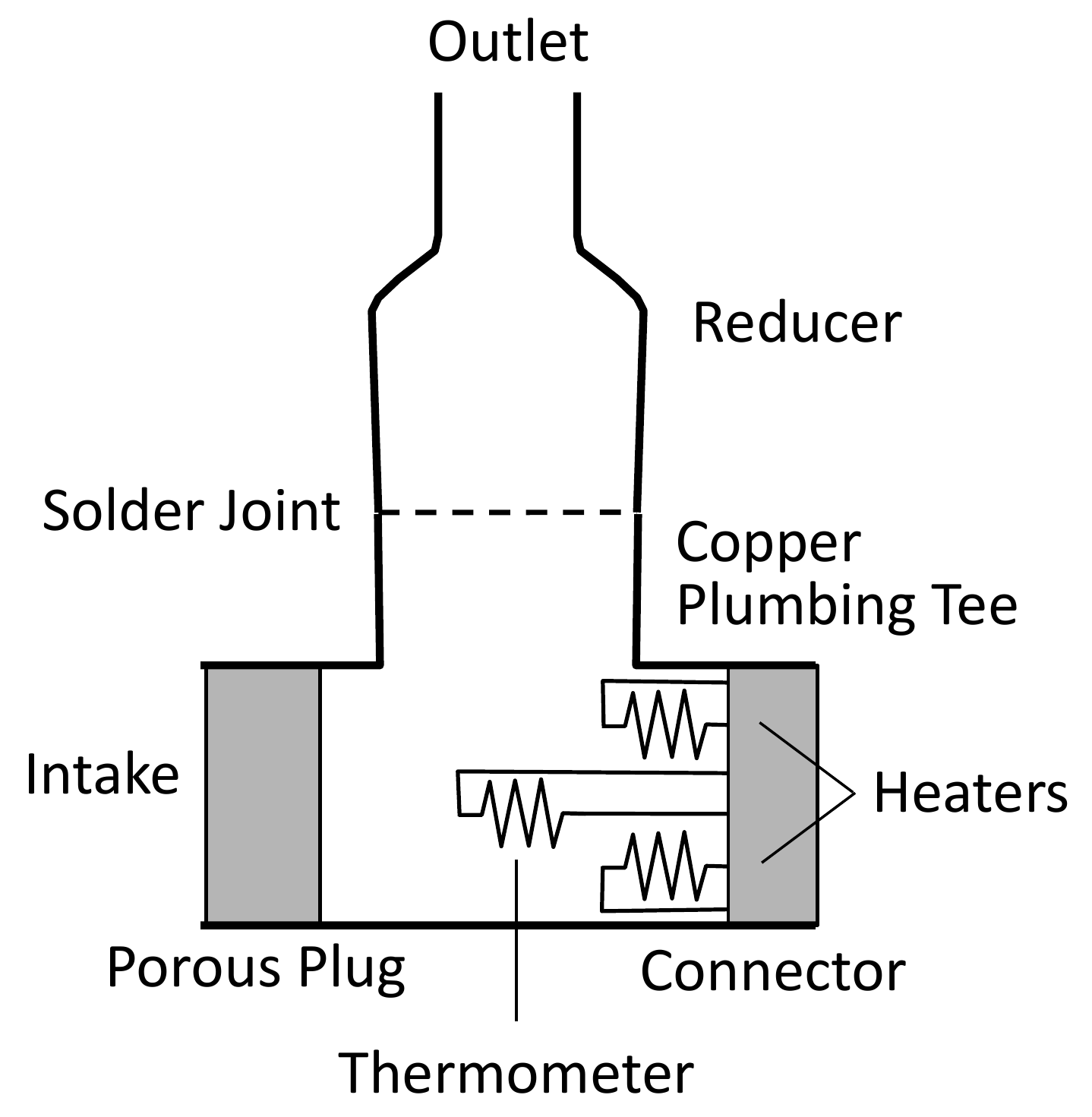} }
\caption{
Major elements of the superfluid pump.
A temperature gradient across the porous plug blocking the pump intake
creates a pressure gradient to force superfluid helium
through the outlet port.
}
\label{pump_schematic}
\end{figure}
%--------------------------------------------------------------------------

\section{Superfluid Pumps}

Figure \ref{pump_schematic} shows the pump design.
A plug with small pores
blocks the intake into the pump body,
which contains a resistive heater.
The fraction of the liquid in the superfluid state
is a function of temperature,
increasing from zero at the $\lambda$ point 
to 100\% at absolute zero.
Raising the temperature of the liquid inside the pump body
decreases the superfluid fraction within the pump.
Osmosis then drives the zero-viscosity superfluid
from the higher concentration in the main bath
across the plug and into the pump body.
The small pore size in the plug 
impedes a corresponding counterflow of the normal component,
creating a pressure differential
\begin{equation}
\Delta P = \rho S \Delta T 
= 57 ~
\left( \frac{T}{1.7~{\rm K}} \right)^{5.5} ~
\left( \frac{\Delta T}{1~{\rm mK}} \right) ~
{\rm Pa} 
\label{london_eq}
\end{equation}
where
$\rho$ is the fluid density,
$S$ is the entropy,
$T$ is the bath temperature,
and $\Delta T$ is the temperature difference
across the plug\cite{london:1938}.
At bath temperature 1.7~K,
a temperature gradient $\Delta T = 50$~mK
will create a pressure 2.9~kPa
sufficient to 
support a LHe column 200~cm
above the bath.
Additional heating
% (or a lower column height)
then provides a steady flow of superfluid helium.

%--------------------------------------------------------------------------
% Figure 4: Pump pix
%--------------------------------------------------------------------------
\begin{figure}[b]
\centerline{
\includegraphics[height=3.5in]{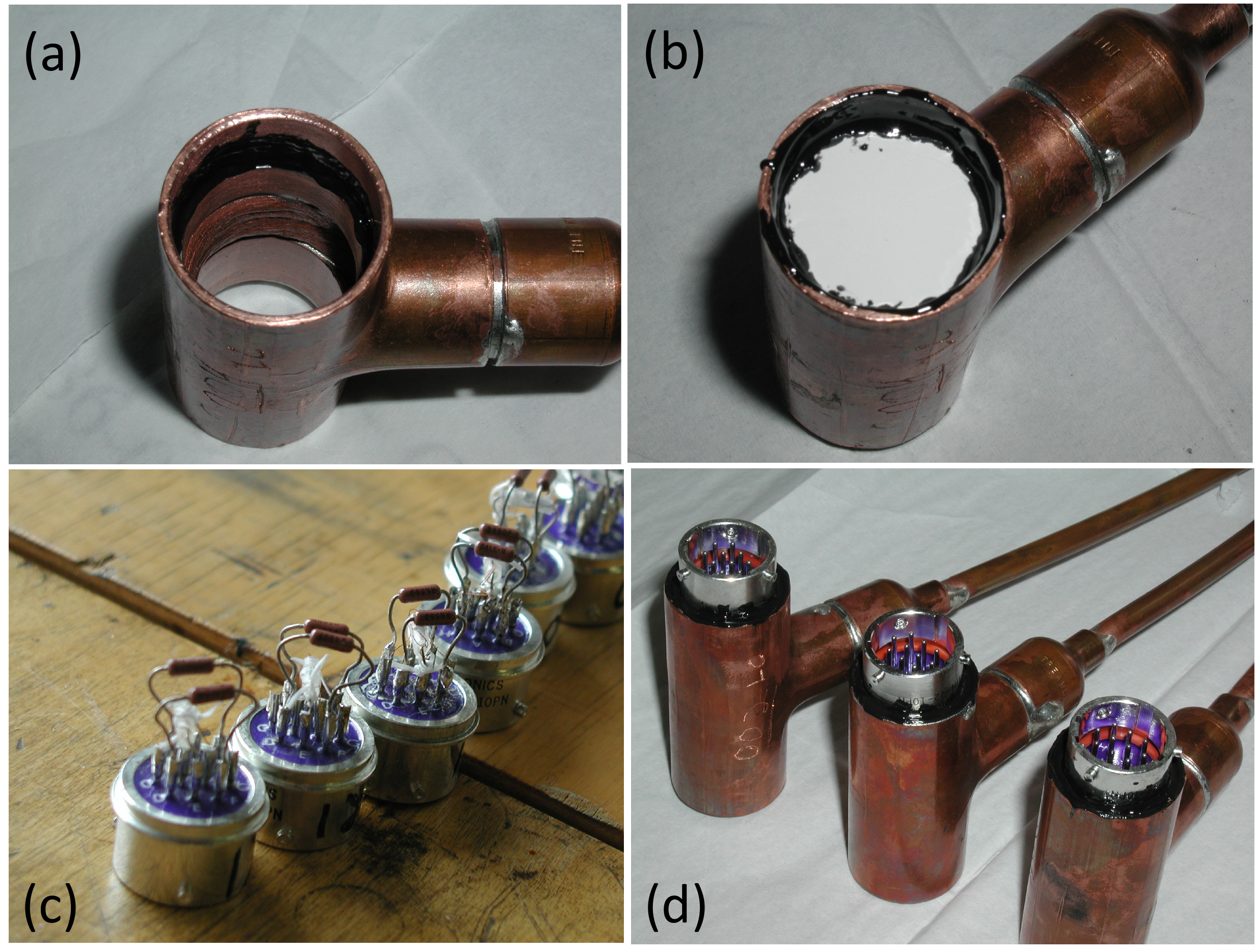} }
\caption{
Superfluid pumps during fabrication.
(a) Copper tee with epoxy for porous plug.
(b) Porous plug in pump intake.
(c) Electrical insert with heater resistors and thermometer.
(d) Assembled pumps showing electrical connector.
}
\label{pump_pix}
\end{figure}
%--------------------------------------------------------------------------

The PIPER superfluid pumps
are designed for ease of manufacture.
Figure \ref{pump_pix} shows the as-built pumps.
Each pump body consists of 
a standard household 25~mm copper plumbing tee.
We first solder a copper reducer
and flared compression fitting
onto the tee ``handle''
which couples to thin-wall stainless steel tubing
on the pump output.
We epoxy a circular disk ground from
CoorsTek P-1/2-BC porous ceramic
into one end of the tee
using Stycast 2850 FT epoxy
to form a porous plug intake 
with pore diameter less than 0.5~$\mu$m.
A circular hermetic connector
epoxied into the opposing end of the tee
provides electrical connectivity.
Two redundant 350~$\Omega$ resistors
wired in parallel
provide ohmic heating within the pump body
while a
resistive ruthenium oxide thermometer
monitors the internal pump temperature.

The electrical power required to produce the desired
temperature (hence pressure) gradient 
depends in part on parasitic heat loss
through the porous plug and the pump body to the bath.
We coat the exterior of the copper pump body
with a 2~mm layer of Stycast 2850 FT epoxy
to provide modest thermal isolation
between the pump interior 
and the main bath.
Although a more sophisticated thermal isolation
would allow the pump to operate at reduced electrical power,
heat flow through the pump body to the bath
supplies part of the heat
dissipated within the bath
to maintain a sufficient efflux of boiloff gas
through the telescope aperture.

%--------------------------------------------------------------------------
% Figure 5: Pump calibration setup
%--------------------------------------------------------------------------
\begin{figure}[b]
\centerline{
\includegraphics[height=3.5in]{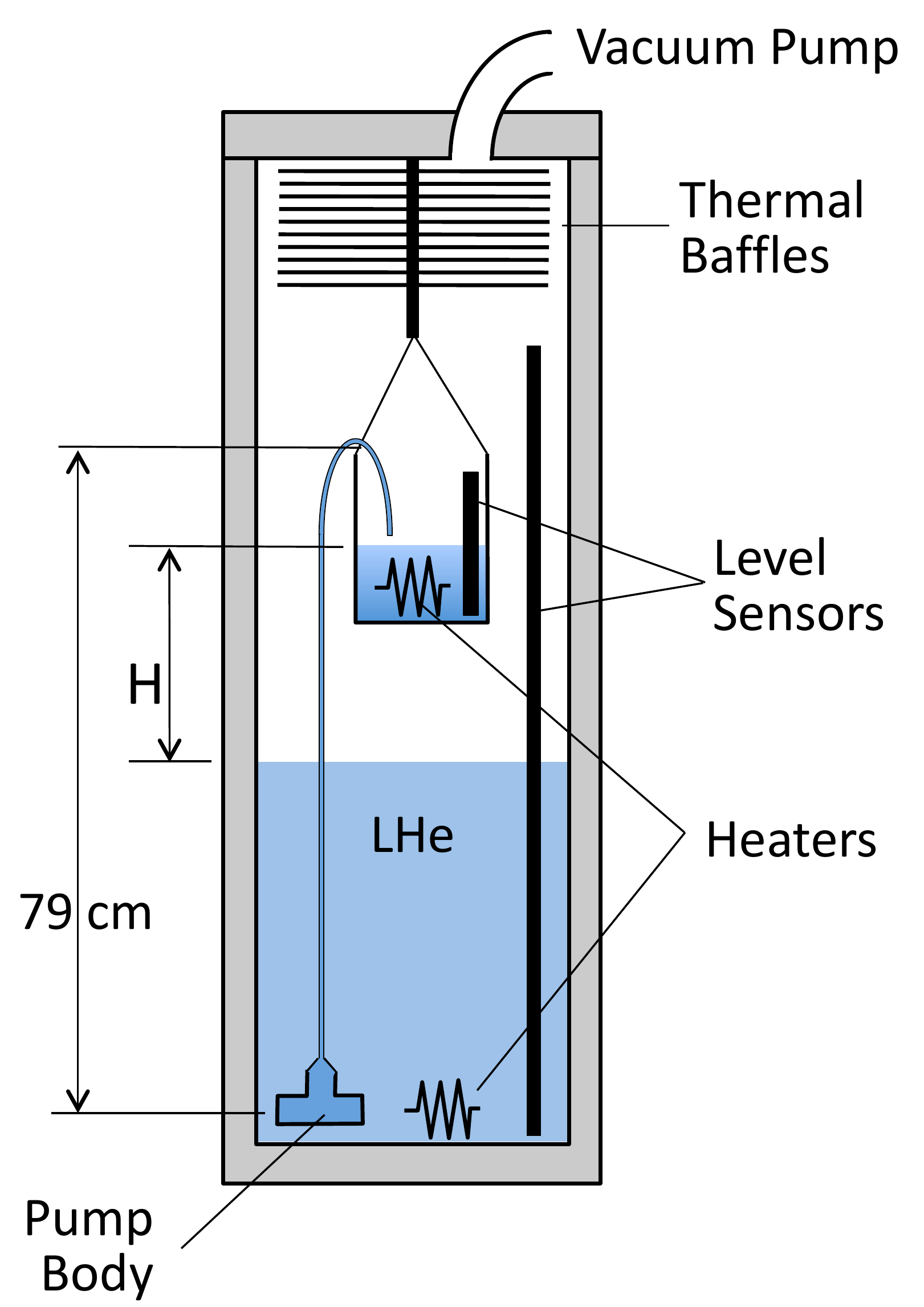} }
\caption{
Test facility for superfluid pump calibration.
The pump outlet is a height $H$ above the liquid level.
}
\label{pump_test_setup}
\end{figure}
%--------------------------------------------------------------------------

We calibrate the as-built performance
for a set of 12 identical pumps
using the test facility shown in Figure \ref{pump_test_setup}.
We mount the pumps
at the bottom of a liquid helium dewar
and pump on the helium space
to maintain the bath at temperatures 1.4--1.7~K
(the temperature range expected during flight).
A thin-wall stainless steel tube
connects the outlet of the pump
to a container positioned above the bath surface.
A thermometer and 
an American Magnetics continuous level sensor
record the temperature and liquid level in the container.
A heater on the bottom of the container
can be powered as needed to remove LHe from the container.

%--------------------------------------------------------------------------
% Figure 6: Measured flow rate
%--------------------------------------------------------------------------
\begin{figure}[b]
\centerline{
\includegraphics[height=3.5in]{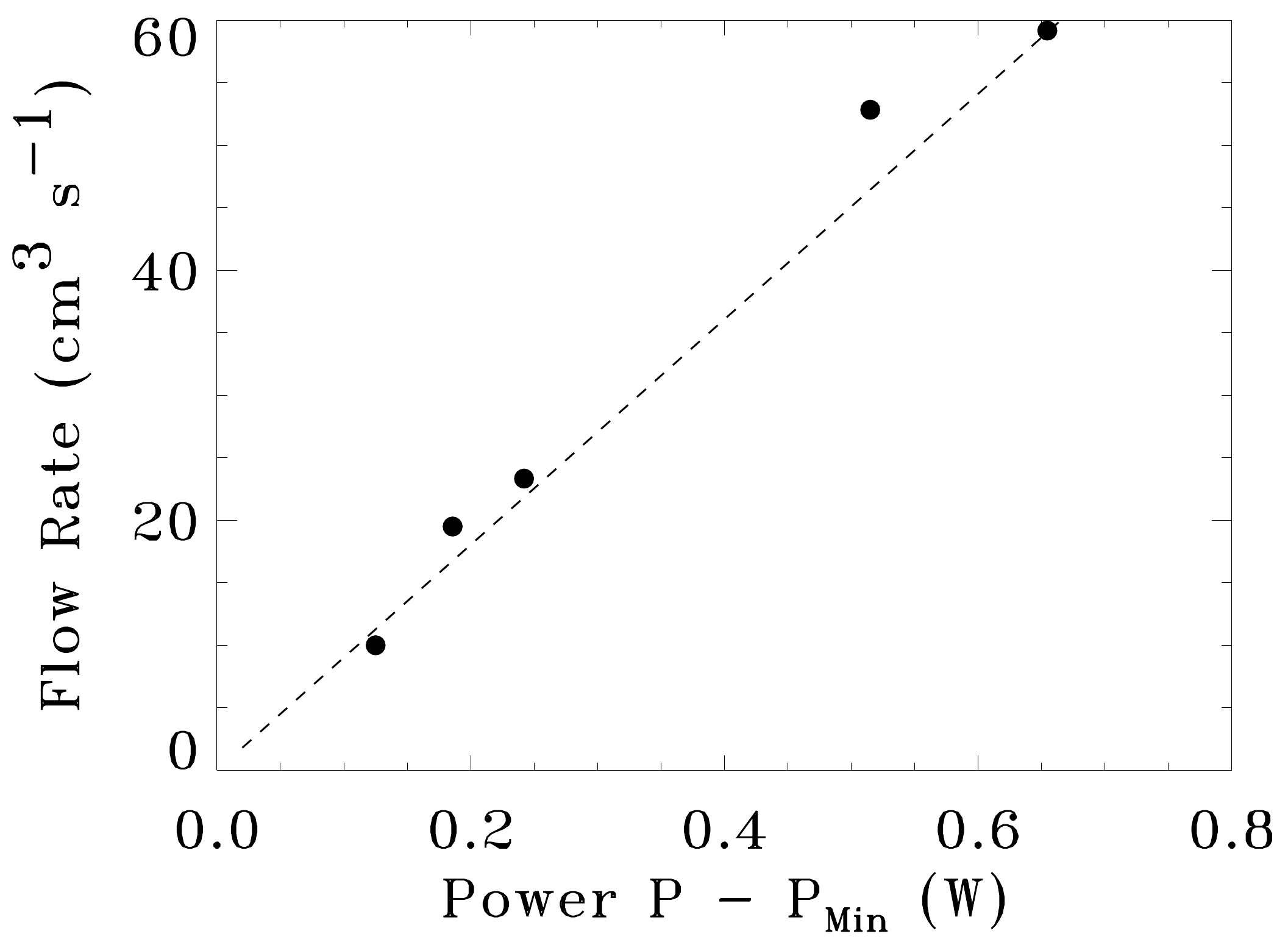} }
\caption{
The flow rate through the pump is 
a linear function of the dissipated heater power.
Filled circles show the measured flow 
for column height $H = 115$~cm
at bath temperature $T = 1.43$~K.
The dashed line is the expected flow
from Eqs. \ref{power_balance} -- \ref{simple_flow_eq}.
}
\label{flow_fig}
\end{figure}
%--------------------------------------------------------------------------

%--------------------------------------------------------------------------
% Figure 7: Relative terms in flow equation
%--------------------------------------------------------------------------
\begin{figure}[b]
\centerline{
\includegraphics[height=5.0in]{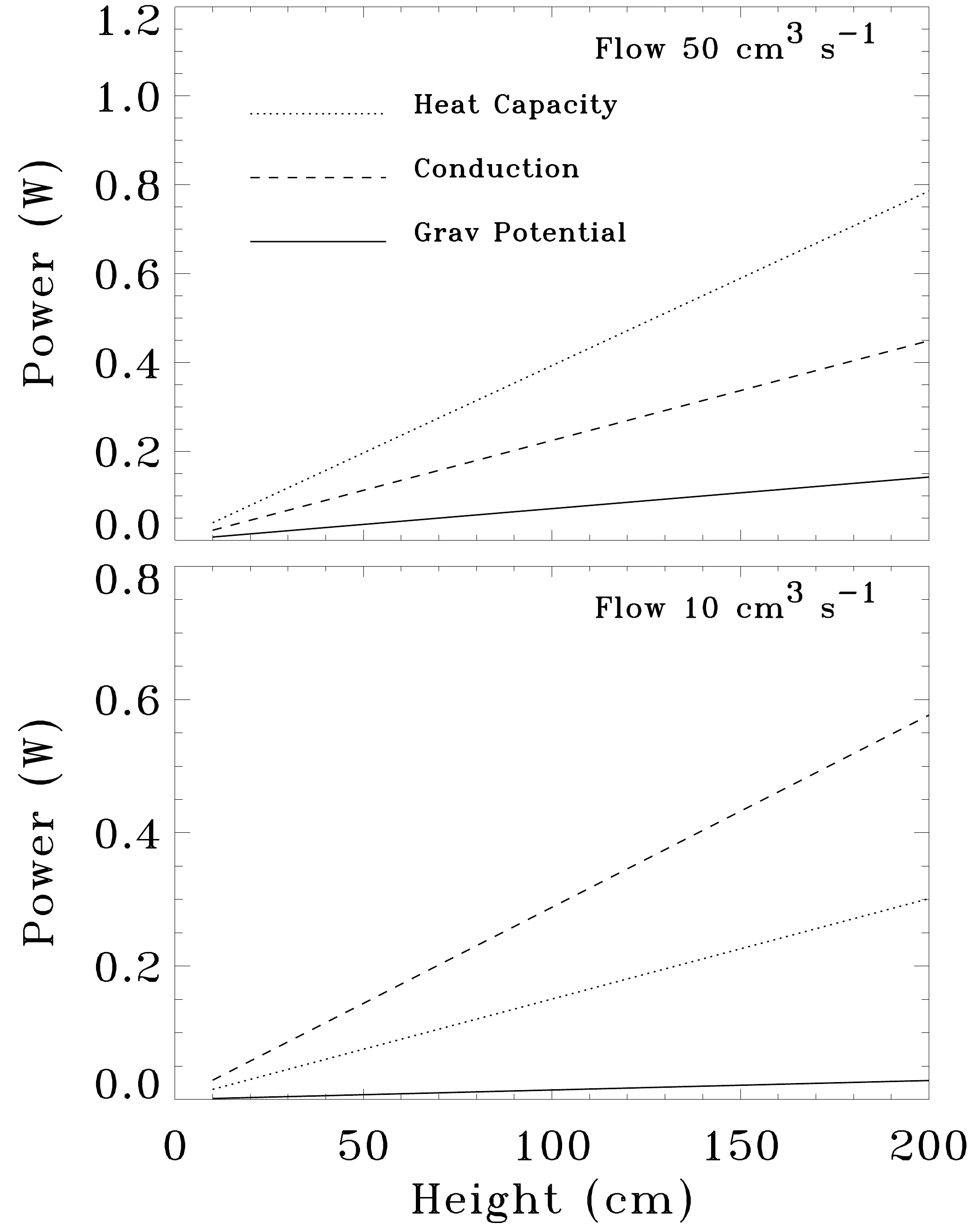} }
\caption{
The pump heater provides power to 
heat the liquid inside the pump (dotted line)
and 
lift the liquid against gravity (solid line).
Parasitic heat flow from the pump interior
back to the bath (dashed line)
becomes increasingly important at low flow rates.
Data are shown for bath temperature 1.7~K.
}
\label{flow_terms}
\end{figure}
%--------------------------------------------------------------------------

The flow rate $F = dV/dt$ of a pump
depends on the electrical power $P_E$ 
dissipated by the pump heater as
\begin{equation}
P_E = G \Delta T
    + Q(F) + W(F) + K(F)
\label{power_balance}
\end{equation}
where
$G$ is the thermal conductivity from the pump interior to the bath,
\begin{equation}
Q = C(T) \Delta T \, \rho \:F
\label{Q_eq}
\end{equation}
is the power required to raise the temperature inside the pump,
\begin{equation}
W = \rho g H \: F
\label{potential_eq}
\end{equation}
is the power required to lift the liquid against gravity,
and
\begin{equation}
K = \frac{ \rho F^2}{2 \pi R^2}
\label{kinetic_eq}
\end{equation}
is the kinetic power.
Here 
$\rho$ is the liquid density,
$T$ is the temperature,
$C$ is the heat capacity,
$g$ is the acceleration from gravity,
$R$ is the radius of the output tubing,
and
$H$ is the difference in height 
between the top of the liquid in the container
vs the top of the liquid in the bath.
With no flow through the pump,
the last three terms in Eq. \ref{power_balance}
are identically zero,
allowing a simple calibration of heat flow
through the pump body.
With the container initially empty,
we increase the voltage to the pump heater
while monitoring the container to determine
the minimum power $P_{\rm min}$
required to raise the helium column to height $H$
without delivering liquid into the container.
Equating
$P_{\rm min} = G \Delta T$,
we find mean value 
$\langle G \rangle = 7.0 \pm 1.4$~W~K$^{-1}$
for the parasitic loss
at $T = 1.43$~K.

%--------------------------------------------------------------------------
% Figure 8: Pump manifold
%--------------------------------------------------------------------------
\begin{figure}[b]
\centerline{
\includegraphics[height=4.0in]{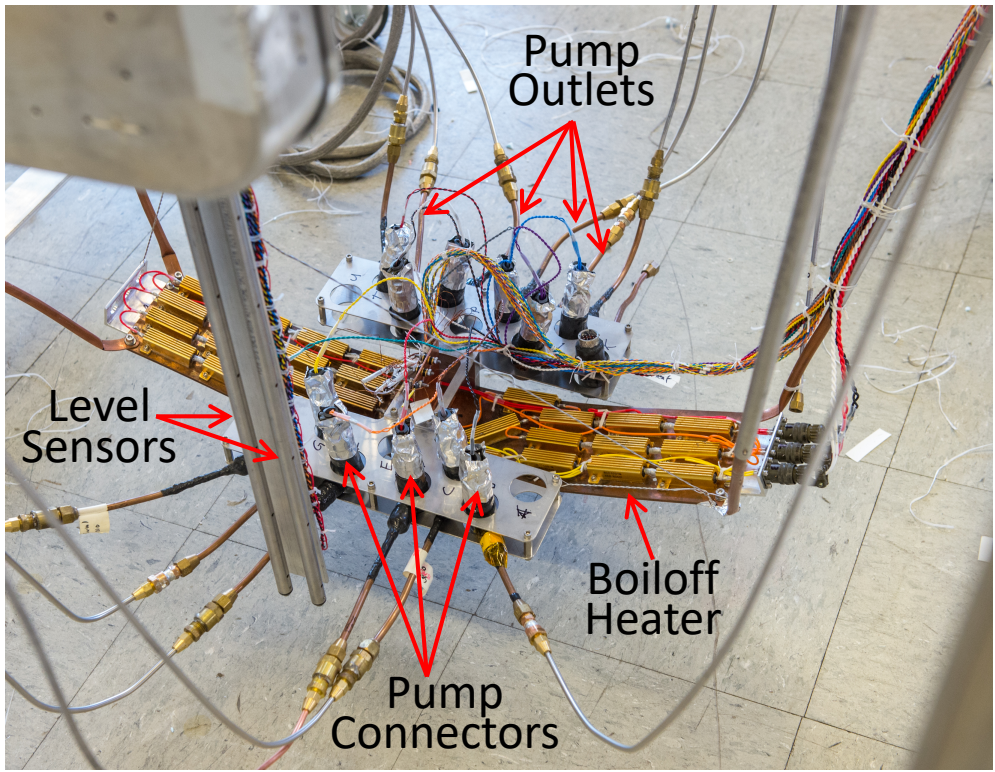} }
\caption{
Set of 12 superfluid pumps in the flight manifold.
}
\label{pump_manifold}
\end{figure}
%--------------------------------------------------------------------------

Increasing the heater power above $P_{\rm min}$
produces a flow 
of liquid helium into the container
\begin{equation}
\frac{dV}{dt} = \frac{1}{\rho} \frac{dm}{dt} 
\propto \frac{P_E - P_{\rm min}}{S~T} ~,
\label{simple_flow_eq}
\end{equation}
where 
$dV/dt$ is the volume change
associated with helium mass flow $dm/dt$
\cite{broulik/hess:1978,
nakai/etal:1996}.
We monitor the liquid level within the container
to determine the change in volume per unit time,
correcting 
the resulting flow rate from the pump
for the passive boiloff within the container
determined by setting $P_E = 0$
and monitoring the container volume
as the LHe boils away.
Figure \ref{flow_fig} shows the measured flow rate
as a function of the heater power,
demonstrating the expected linear dependence on the heater power,
\begin{equation}
\frac{dV}{dt} = \alpha (P_E - P_{\rm min})
\label{fitted_flow_eq}
\end{equation}
with coefficient
\begin{equation}
\alpha = 95 \pm 7 ~\left( \frac{T}{1.43~{\rm K}} \right)^{-6.5}
~ {\rm cm}^{3}~{\rm s}^{-1} ~{\rm W}^{-1}
\label{alpha_eq}
\end{equation}
Figure \ref{flow_terms} shows the relative amplitudes
of the terms in Eq. \ref{power_balance}
at temperature 1.7~K.
The bulk of the heater power goes to 
heating the liquid as it enters the pump
and the parasitic heat flow through the pump body back to the bath.
The power required to heat the liquid depends on the flow rate
and dominates at higher flow rates (top panel),
while the parasitic loss through the pump body
is nearly independent of flow rate
and becomes more important at low rates.
The power needed to raise the liquid against gravity
is always sub-dominant,
while the kinetic term $K$ is negligible.
At temperatures 1.5--1.7 K
appropriate for float altitude 30--35~km,
applied heater power of $P_E \sim 1$~W
induces flow rates for a single pump 
of 15~cm$^3$~s$^{-1}$
to a height 2 meters above the bath.

The linear dependence 
shown in Fig. \ref{flow_fig} does not extend to arbitrarily high flow rate.
Above a critical velocity,
turbulence sets in
so that additional heater power
produces little additional flow
\cite{broulik/hess:1978,
nakai/etal:1996}.
The demonstrated peak flow of 
60~cm$^3$ s$^{-1}$
(215 liters~hr$^{-1}$)
for a single pump
is sufficient to maintain the PIPER optics at the bath temperature;
consequently,
we do not attempt to determine the maximum flow rate
for the PIPER pump configuration.

% -------------- Table 1: Pump Allocation --------------
\begin{table}[b]
{
\small
\caption{Pump Allocation}
\label{pump_table}
\begin{center}
\begin{tabular}{l l c}
\hline 
Pump	&	Outlet Target	& Outlet Height (cm) \\
\hline
1	&	Port Dewar Aperture		& 200	\\
2	&	Port Secondary Mirror		& 130	\\
3	&	Port Primary Mirror		& 180	\\
4	&	Starboard Primary Mirror	& 180	\\
5	&	Starboard Secondary Mirror	& 130	\\
6	&	Starboard Dewar Aperture	& 200	\\
7	& 	Port VPM			& 180	\\
8	&	CADR Feedthrough		& 120	\\
9	&	CADR Feedthrough		& 120	\\
10	&	Receiver Heat Sink		& 100	\\
11	&	Receiver Heat Sink		& 100	\\
12	&	Starboard VPM			& 180	\\	
\hline
\end{tabular}
\end{center}
}
\end{table}
%------------------------------------------------------------

\section{Flight Performance}

PIPER uses a set of 12 identical superfluid pumps 
to maintain the reflective optics
at temperatures below 3~K.
Figure \ref{pump_manifold}
shows the set of pumps mounted in the flight manifold.
A metal frame holds the pumps near the dewar bottom,
oriented so the pump inlet port is horizontal.
We direct the output of individual pumps
to the rear surfaces of the 
VPM, primary mirror, and secondary mirror
for each of the two telescopes.
An additional pump for each telescope
directs superfluid helium onto the top of the dewar wall,
at the point where the telescope beam exits the dewar aperture.
Two redundant pumps deliver LHe
onto the vacuum feedthrough
housing the high-current wiring for the 
continuous adiabatic demagnetization refrigerator (CADR).
The final two pumps deliver LHe to a small container
mounted atop the receiver cryostat,
providing a stable heat sink
for the receiver
even when the LHe level within the dewar
drops below the receiver
(Fig \ref{piper_schematic}).
Table 1 summarizes the pump allocation
and the height of each outlet above the dewar floor.

\begin{figure}[b]
\centerline{
\includegraphics[height=4.0in]{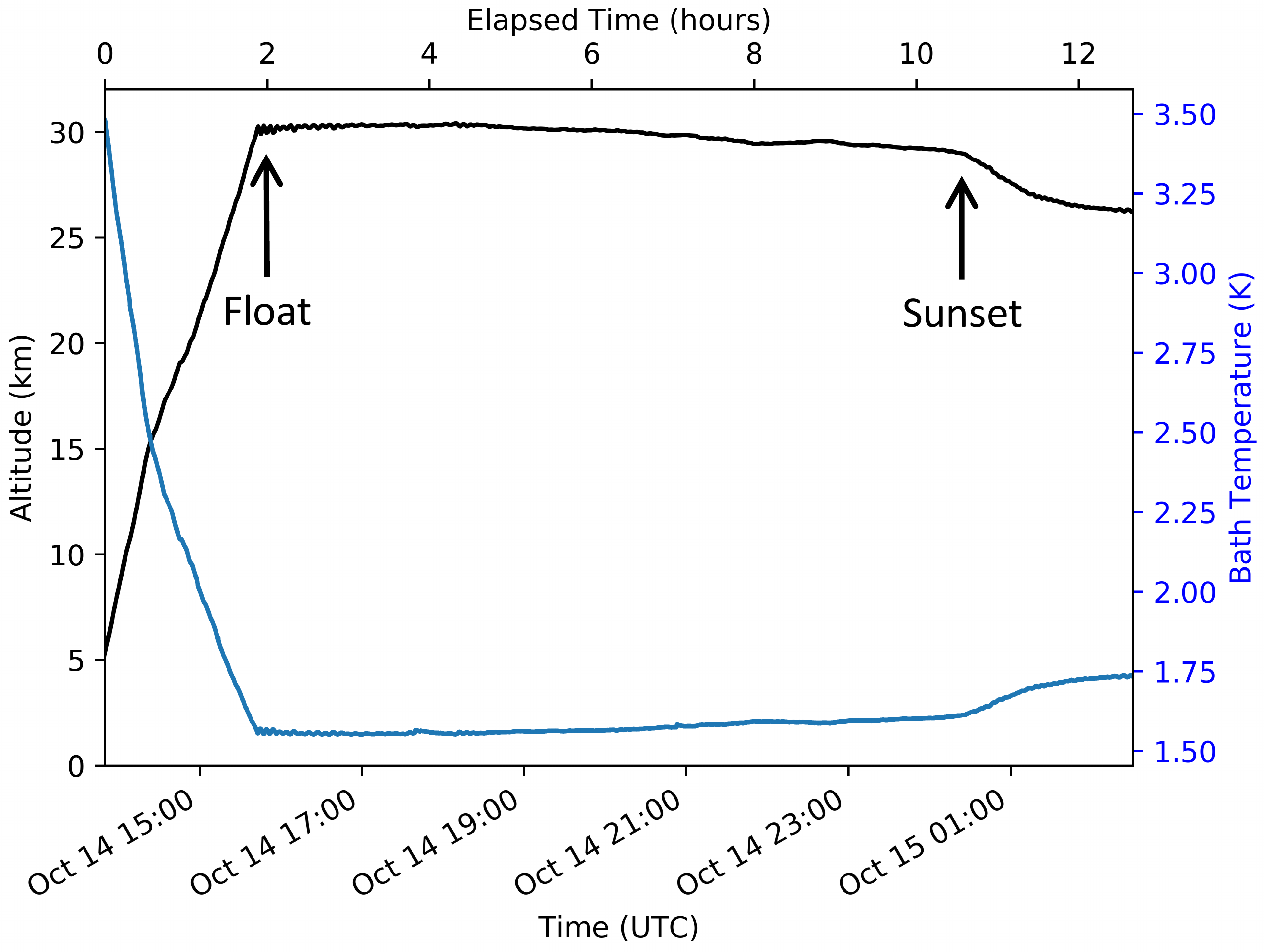} }
\caption{
Payload altitude and bath temperature profile for the 2019 flight.
}
\label{temp_vs_alt}
\end{figure}

PIPER has flown twice:
an engineering flight in 2017
and a science flight in 2019.
Both flights demonstrated similar cryogenic performance.
For simplicity, we show performance from the
2019 Oct 14 flight 
which launched at 13:36 UTC 
from Ft Sumner, NM
with 1515 liters (195~kg) of liquid helium in the bucket dewar.
By 15:46 the payload reached float altitude of 29.9 km.
Figure \ref{temp_vs_alt}
shows the temperature and altitude profile
for this flight.
Dual vents in the protective cover allow the dewar to vent during ascent
so that the bath temperature is determined by the ambient pressure,
reaching a stable value of 1.7~K at float altitude.
The continuous LHe level sensor was turned on at 16:10,
showing 985 liters (143~kg) remaining after ascent.
The 35\% drop in LHe volume during ascent
corresponds to a 27\% mass loss
after accounting for the higher LHe density at 1.7 K,
and is consistent with 
a 25\% LHe mass loss during the superfluid transition
plus minor losses
from parasitic heat leaks during ascent.
Cooling of the balloon after local sunset 
causes a loss in buoyancy;
the resulting drop in altitude
is reflected in modestly higher bath temperatures.

%--------------------------------------------------------------------------
% Figure 9: Altitude and bath temperature profile
%--------------------------------------------------------------------------
%--------------------------------------------------------------------------

%--------------------------------------------------------------------------
% Figure 10: Optics pump and temps at initial cooldown
%--------------------------------------------------------------------------
\begin{figure}[b]
\centerline{
\includegraphics[height=4.5in]{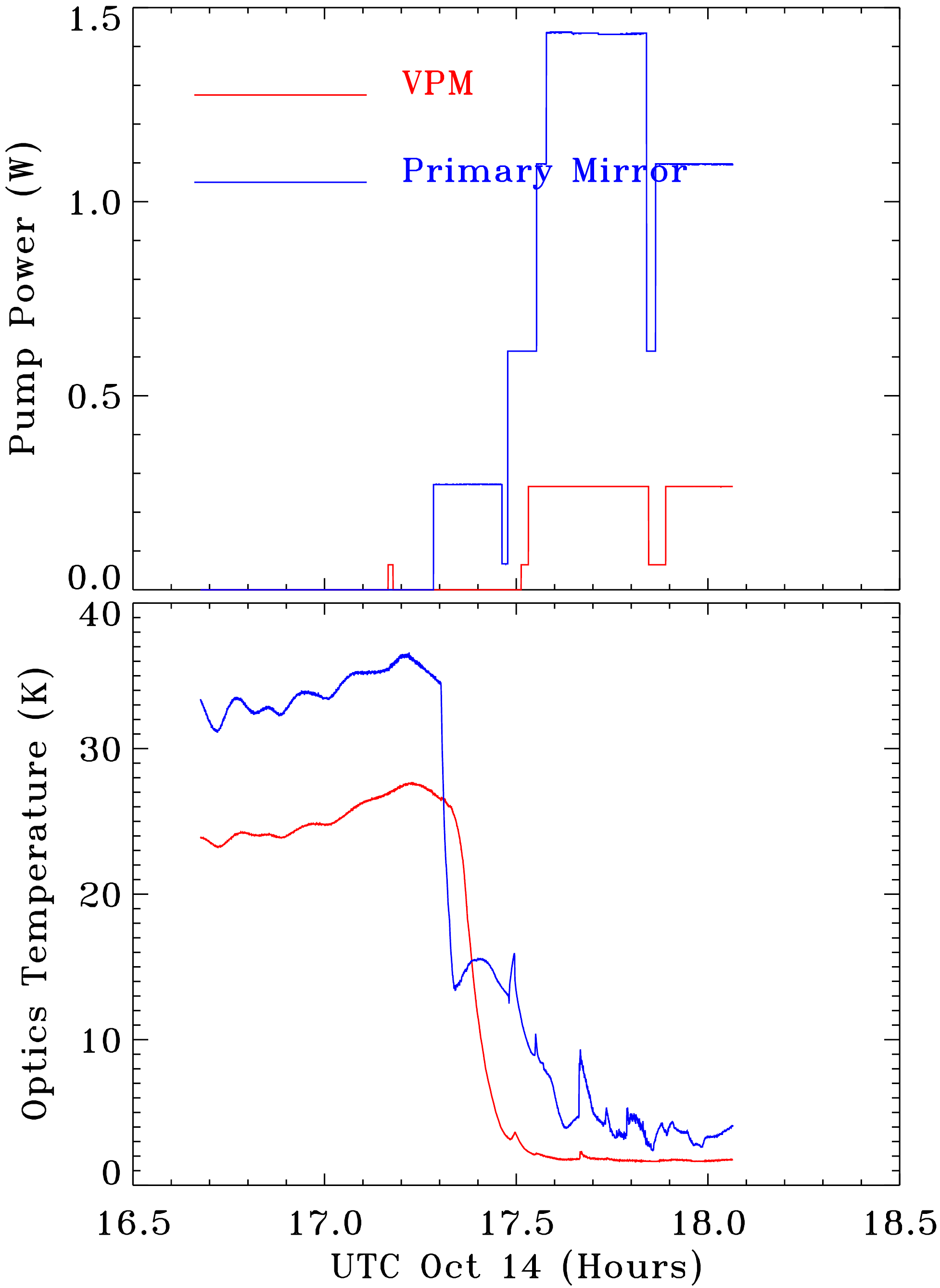} }
\caption{
Superfluid pump power (top pane)
and optics temperature (bottom panel)
for the intial cooldown of the starboard telescope at float.
Applying power to the superfluid pumps
cools the VPM and primary mirror
mounted 200~cm above the dewar floor
to temperatures near the LHe bath.
}
\label{optics_cooldown}
\end{figure}
%--------------------------------------------------------------------------

At 17:15 the superfluid pumps were powered
to cool the optics.
Figure \ref{optics_cooldown} shows the optics temperature
and commanded pump power
for the two optical surfaces mounted highest within the dewar
(the VPM and primary mirror).
Prior to initiating LHe flow through the pumps,
both surfaces showed a slow warming trend
as temperatures stratified within the dewar.
At 17:17 UTC the pump servicing the starboard primary mirror
was turned on,
followed promptly by a sharp drop in mirror temperature
from 35~K to 3~K.
Within 200 seconds,
the starboard VPM temperature also began to fall,
although the VPM pump was not commanded on until 17:31.
Since the VPM and primary mirror are mounted 
on opposite sides of the dewar
(Fig \ref{piper_schematic}),
the probable cause for the observed cooling
is increased gas flow 
as LHe boils on warm elements within the dewar.
Once the optics cooled, 
they remained near the bath temperature
for the remainder of the flight.

%--------------------------------------------------------------------------
% Figure 11: Aperture pump and temps 
%--------------------------------------------------------------------------
\begin{figure}[b]
\centerline{
\includegraphics[height=4.5in]{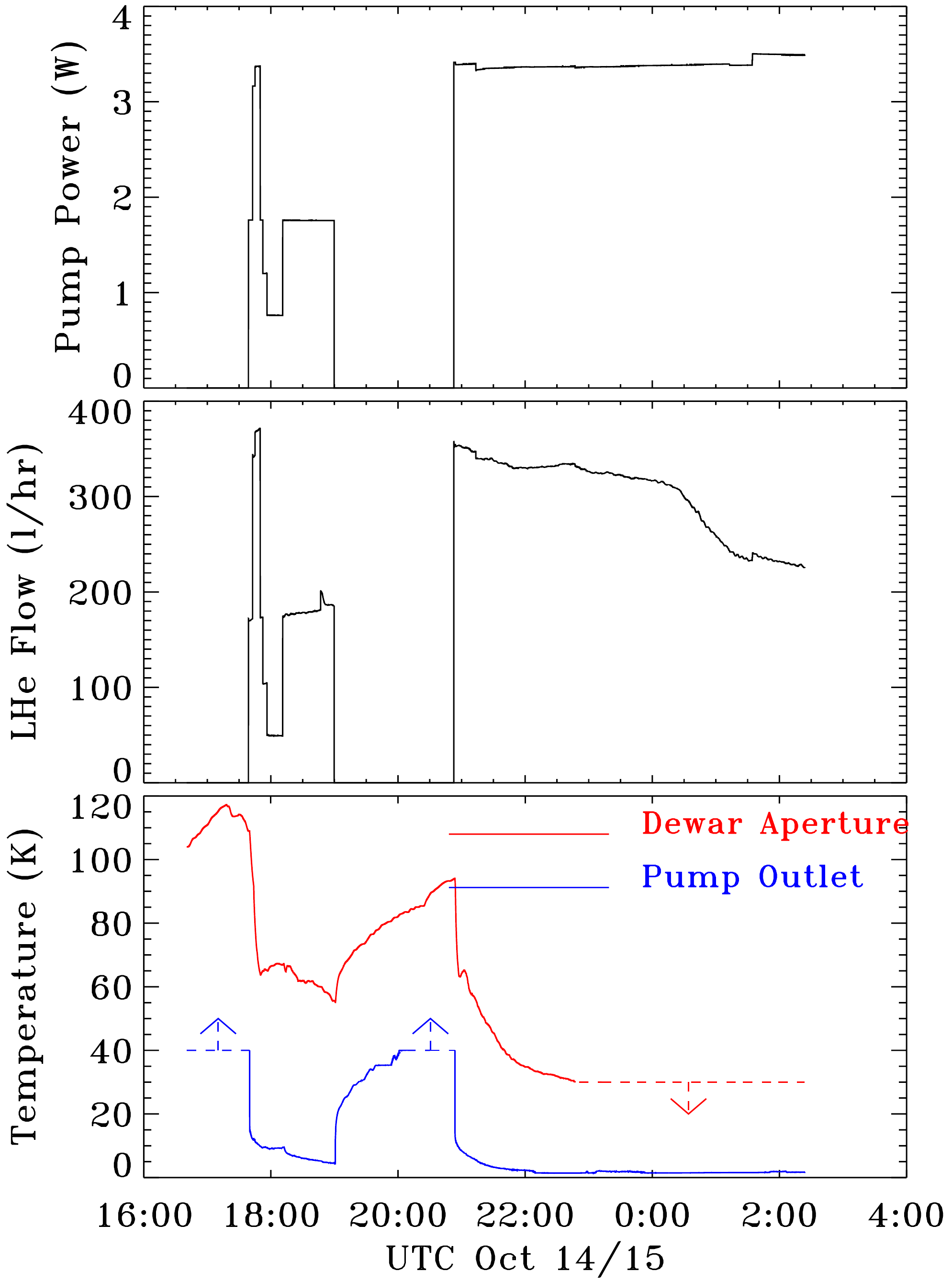} }
\caption{
Superfluid pump power (top panel),
LHe flow rate (middle panel)
and component temperatures (bottom panel)
for the pump servicing the dewar starboard aperture.
A single superfluid pump maintains the top of the dewar wall
at temperatures below 30~K.
}
\label{aperture_cooldown}
\end{figure}
%--------------------------------------------------------------------------

The power loading from the sky to the detectors is small
(below 1~pW).
Even modest beam spillover onto warm surfaces
could saturate the detectors.
The warmest portion of the dewar interior
is the top of the wall at the dewar aperture.
To minimize optical power from the dewar aperture,
we cool the aperture using a pair of pumps
(one for each telescope).
Figure \ref{aperture_cooldown}
shows temperatures and commanded pump power
for the starboard dewar aperture.
A ruthenium oxide thermometer monitors the temperature
at the pump outlet,
but saturates at temperatures above 40~K.
A diode thermometer monitors the temperature 
at the top of the dewar wall
some 10~cm above the pump outlet,
but saturates at temperatures below 30~K.
At 17:39 the starboard aperture pump was powered on 
for 21 minutes
at power levels 0.8 to 3.3~W,
corresponding to commanded flow rates 50 to 350 liters~hr$^{-1}$.
Temperatures at the dewar aperture
promptly fell from 120~K to 60~K.
Between 19:00 and 20:52 the aperture pumps were turned off,
resulting in a slow rise in aperture temperatures
from 60~K to 90~K.
At 20:53 the aperture pump was turned on again
for a commanded flow rate of 350 liters~hr$^{-1}$,
after which the aperture temperature fell below 30~K
where it remained for the duration of the flight.

%--------------------------------------------------------------------------
% Figure 12: Dewar stratification
%--------------------------------------------------------------------------
\begin{figure}[b]
\centerline{
\includegraphics[height=4.5in]{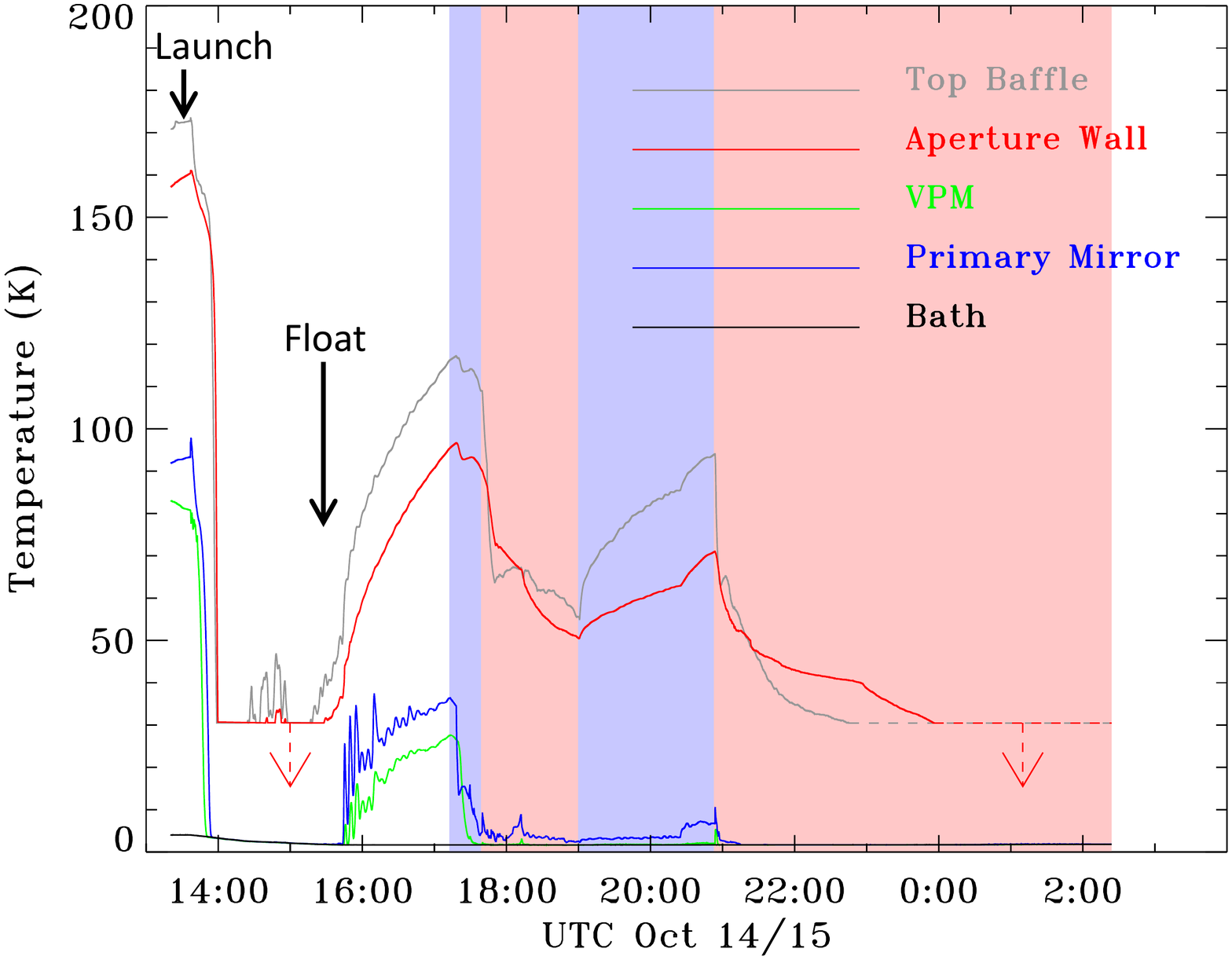} }
\caption{
Temperatures within the dewar depend on the boilof rate
and pump operation.
During ascent, helium boiloff alone is sufficient
to maintain the entire interior near the bath temperature.
Once float altitude is reached,
temperatures within the dewar stratify
until the pumps are turned on.
The blue bands
indicate periods when pumps cooled the optics
but not the dewar aperture.
Pink bands indicate periods
when the aperture pumps were also turned on.
Thermometers for the top baffle and aperture wall
are unable to record temperatures below 30~K,
and are shown as upper limits for these times.
}
\label{dewar_gradient}
\end{figure}
%--------------------------------------------------------------------------

Note that cold gas efflux from the passive helium boiloff rate
is not normally sufficient by itself to cool the PIPER optics
or prevent warming in the upper portion of the dewar.
Figure \ref{dewar_gradient} shows the temperature profile
for selected elements in the dewar throughout the flight.
Prior to launch, 
temperatures within the dewar are stratified,
with components further above the bath
at progressively higher temperatures.
During ascent,
the rapid helium boiloff
cools the telescope and dewar interior
to the bath temperature.
Once float altitude is reached,
the boiloff rate decreases dramatically
and the dewar re-stratifies.
Direct application of superfluid LHe
onto the telescope optics
cools the optics to the bath temperature,
but is not in itself sufficient to cool the entire dewar.
Spraying LHe onto the dewar top wall
restores near-isothermal operation
at the bath temperature.

\section{Discussion}

PIPER uses a set of 12 identical superfluid pumps
to maintain the optical surfaces at 1.7~K.
Each pump is capable of producing flow rates 
in excess of 200 liters~hr$^{-1}$.
Monitoring the liquid level within the dewar
despite the resulting LHe spray
and
high boiloff gas flow
proved challenging.
The continuous level sensors 
flow a 75~mA current through a superconducting filament.
In normal operation,
the portion of the filament
submerged in LHe
remains superconducting,
while ohmic heating drives the portion above the liquid
into the normal state.
The electrical resistance across the entire filament
is thus proportional
to the segment length above the liquid.
Excessive cooling of the portion above the liquid,
either by gas flow
or liquid splashes, 
maintains the filament in the superconducting state
to produce erroneous readings.
We minimize this effect 
by mounting the level sensors within a hollow metal tube
to shield it from both gas flow and liquid splashes.
Small holes drilled through the tube every 10~cm in height
allow the liquid level inside to equilibrate with the main dewar tank.
With this modification,
the level sensors performed as designed throughout flight.

%--------------------------------------------------------------------------
% Figure 13: Helium volume vs time 
%--------------------------------------------------------------------------
\begin{figure}[b]
\centerline{
\includegraphics[height=4.5in]{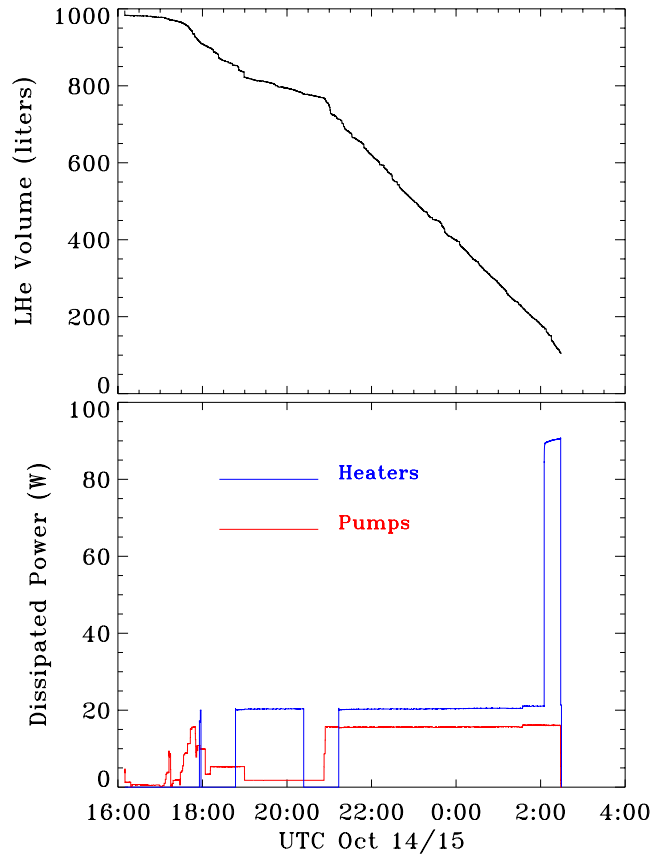} }
\caption{
The liquid helium volume measured during float operation
(top panel)
can be compared to the electrical power dissipated by
the superfluid pumps and boiloff heaters (bottom panel).
The LHe loss rate
is dominated by flow from the pumps
onto warm portions of the dewar,
with only a minor contribution from dissipated heater power.
}
\label{boiloff_rate_fig}
\end{figure}
%--------------------------------------------------------------------------

PIPER's open-aperture operation requires a constant efflux
of helium gas to prevent condensation of atmospheric gases
onto the cold optics.
The helium boiloff rate
depends on both the power dissipated within the dewar
as well as evaporation from the liquid flow 
directed onto warm portions of the dewar.
Figure \ref{boiloff_rate_fig} 
shows the LHe volume within the dewar
throughout operation at float altitude.
Prior to initiation of pump activity,
the observed boiloff rate of 6.4 liters~hr$^{-1}$
is consistent with the typical quiescent loss
8--13~liters~hr$^{-1}$
measured during ground tests.
During the initial pump checkout
from 17:40 to 18:45 UTC,
the boiloff rate increased to
96~liters~hr$^{-1}$.
From 18:45 to 20:20
a boiloff heater on the dewar bottom was powered at 20.3~W
while all pumps except the two VPM pumps were turned off.
Boiloff for this period fell to 29~liters~hr$^{-1}$.
At 20:55
the pumps servicing the primary mirrors, VPMs, and dewar aperture ports
were turned on,
followed by the boiloff heater at 21:14.
The boiloff rate then increased to 110~liters~hr$^{-1}$.
Finally, at 02:05 Oct 15
a larger boiloff heater was powered at 90~W
to dissipate remaining cryogen
in preparation for flight termination and descent.
The boiloff rate then increased to 177~liters~hr$^{-1}$.

%--------------------------------------------------------------------------
% Figure 14: Excess boiloff vs aperturer pump flow
%--------------------------------------------------------------------------
\begin{figure}[b]
\centerline{
\includegraphics[height=4.5in]{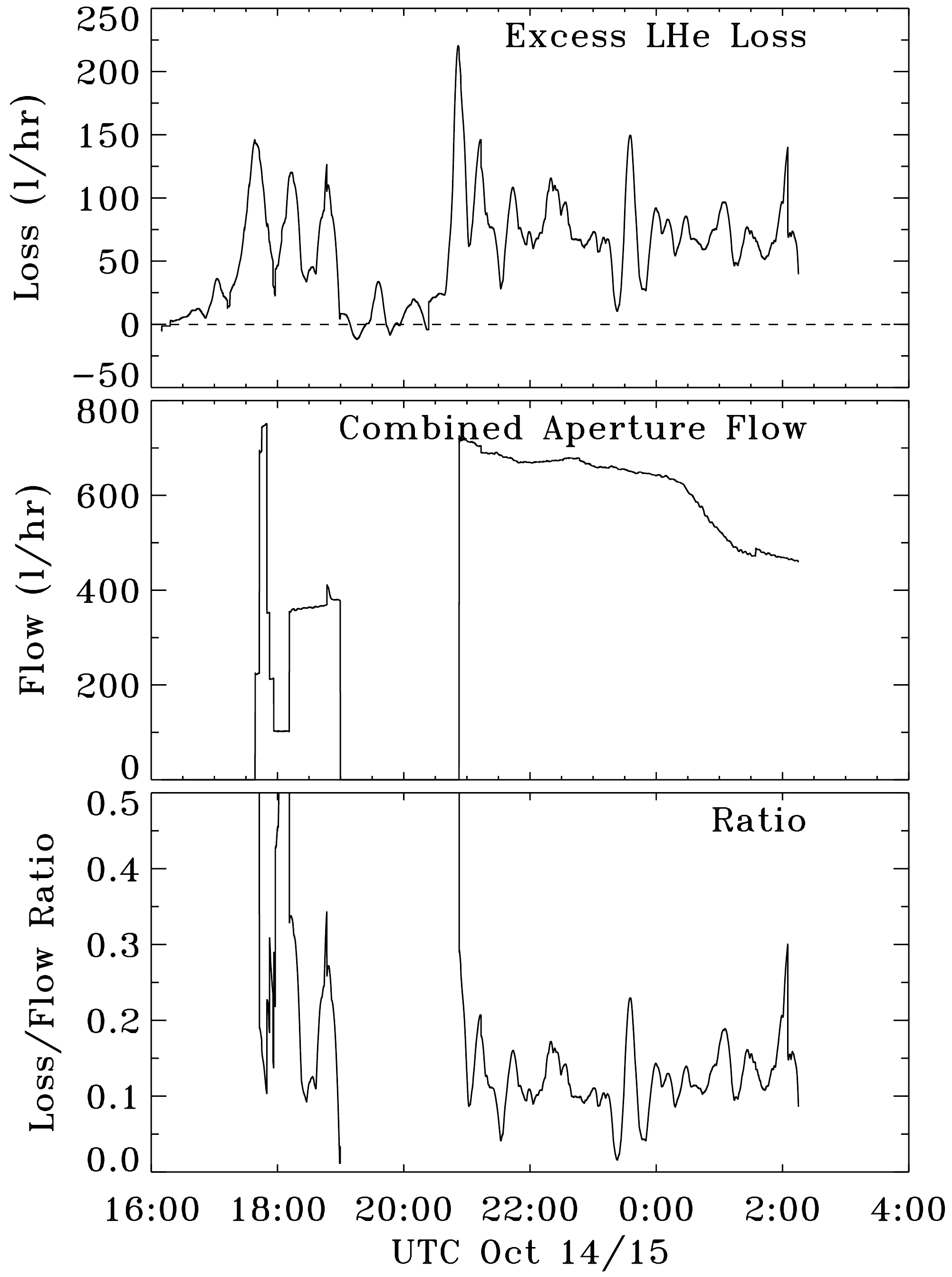} }
\caption{
Helium loss in excess of boiloff from heater power dissipation (top panel)
can be explained by the flow of LHe onto warm portions of the dewar
(middle panel).
The decrease in flow rate to the dewar aperture 
after 22:00
results from the change in column height $H$
as the liquid level falls within the dewar
and the higher bath temperature after sunset.
After the initial cooldown near 18:00 UTC,
the observed excess is consistent with a 15\% loss
from the LHe flow to the dewar aperture (bottom panel).
}
\label{excess_loss}
\end{figure}
%--------------------------------------------------------------------------

The measured helium loss
exceeds the minimum value derived from the
electrical power dissipated within the liquid
by the superfluid pumps and boiloff heaters.
We ascribe the excess loss
to the flow of liquid onto warmer components
near the top of the dewar.
Figure \ref{excess_loss}
compares the excess loss rate
to the commanded flow  
for the combined
port and starboard aperture pumps.
We correct the measured LHe volume loss rate
for the expected loss from the electrical power dissipation,
\begin{equation}
\left. \frac{dV}{dt} \right\rvert_{\rm excess} = 
\left. \frac{dV}{dt} \right\rvert_{\rm obs}
  - ~
  \frac{P_E + P_H}{\rho ~ L_V}
\label{excess_loss_eq}
\end{equation}
where
$P_E$ is the pump electrical power,
$P_H$ is the boiloff heater power,
$\rho$ is the LHe density,
and
$L_V$ is the latent heat of evaporation.
When the pumps are off,
the excess loss is consistent with zero, as expected.
We then compare the excess loss to the predicted flow rate
(Eq. \ref{fitted_flow_eq})
delivered to the top of the dewar
by the port and starboard aperture pumps combined.
The excess helium loss
can be explained 
if roughly 15\% of the helium
sprayed onto the dewar aperture
evaporates to cool the metal top of the dewar,
with the remaining 85\% flowing back into the dewar.
If correct,
this interpretation suggests
that there is little penalty in helium loss
incurred by "over-pumping" liquid to the top of the dewar
in excess of the minimum flow required to cool the aperture.

\section{Conclusions}

We describe a simple method
to construct superfluid liquid helium pumps
for the PIPER balloon-borne telescopes.
Electrical power of 3~W supplied to heater resistors within each pump
deliver a fluid flow as much as 
100~cm$^3$~s$^{-1}$
(360~liters~hr$^{-1}$)
to heights 200 cm above the helium bath.
The resulting flow of superfluid LHe
directed onto the telescope optics
maintains the entire telescope
at the 1.7~K bath temperature.
Comparison of liquid loss within the bath
to the total dissipated heater power
suggests that the bulk of the helium flow
directed onto the optics and the top of the dewar wall during flight
simply recirculates back to the bath.
If launched with the maximum practical helium load
of 2600 liters,
the observed boiloff rate of 110~liters~hr$^{-1}$
would allow 19 hours or more of 
fully cryogenic observations at float altitude.

% Add acknowledgments for Paul Cursey, CSBF staff, etc
\begin{acknowledgments}
We thank the CSBF staff for launch support
and 
gratefully acknowledge 
N. Bellis,
P. Cursey,
N. Gandilo,
S. Pawlyk,
and
P. Taraschi
for their contributions to the PIPER flight.
Support for development of the superfluid pumps was provided by 
NASA WBS 51-188-02-54.
PIPER is supported by NASA WBS 399131.02.06.04.08. 
\end{acknowledgments}

% Data availability
\vspace{3mm}
{\bf
{\small
DATA AVAILABILITY STATEMENT}} 

The data that support the findings of this study 
are available from the corresponding author upon reasonable request.

% --------------------- References ---------------------
\bibliography{piper_cryo_rsi}		% File name piper_cryo_rsi.bib

% That's all there is, kiddies.  Ride off into the sunset!
\end{document}